\documentclass[aip,jmp,amsfonts,showpacs,preprint]{revtex4}
\usepackage{amssymb}
\usepackage{mathrsfs}
\begin{document}
\title{Bounds for Fisher information and its production under flow}
\author{{\sc Takuya Yamano}}
\email[Email: ]{yamano@amy.hi-ho.ne.jp}
\affiliation{ 
Department of Information Sciences, Faculty of Science, Kanagawa University, 
2946, 6-233 Tsuchiya, Hiratsuka, Kanagawa 259-1293, Japan}

\begin{abstract}
We prove that two well-known measures of information are interrelated in interesting and useful 
ways when applied to nonequilibrium circumstances. A nontrivial form of the lower bound for the 
Fisher information measure is derived in presence of a flux vector, which satisfies the continuity 
equation. We also establish a novel upper bound on the time derivative (production) in terms of the 
arrow of time and derive a lower bound by the logarithmic Sobolev inequality. These serve as the 
revealing dynamics of the information content and its limitations pertaining to nonequilibrium processes. 
\medskip
\end{abstract}
\maketitle

\section{Introduction}
A fundamental aspect of physical systems being out of equilibrium lies in the existence of 
limits in information measures that the systems possess. It is known that a few measures of 
information only determined by probability distributions are closely related, but studies 
on their bounds had not been done until relatively recently when the flow vectors exist.  
The Fisher information measure, hereafter denoted as $I$, is an intriguing measure behind 
physical laws \cite{Frieden} and it appears as the basic ingredient in bounding  
entropy production. Specifically, the time derivative of the Shannon entropy $dS/dt$ is 
bounded by $I$ \cite{Niko,Brody}. In this sense, there is a fundamental motivation in probing 
the interplay between physical entities and information. Fisher information is defined using the 
time-dependent probability distribution $f(\vec{x},t)$ as   
\begin{eqnarray}
I(t)=\left\langle \left(\frac{\nabla f}{f}\right)^2\right\rangle_f = \sum_{k=1}^d
\left\| \frac{\nabla_k f}{\sqrt{f}}\right\|_2^2,\label{eqn:Idef}
\end{eqnarray}
where $\langle \cdot\rangle_f$ denotes the average with $f$ over the domain of 
phase space coordinate $\vec{x}$, and the $\| \cdot\|_2$ is $L^2$ norm. In the second expression, 
we use $\nabla_k=\partial/\partial x_k, (k=1,\ldots, d)$ with dimension $d$.
At the moment of research, two different types of lower bounds are obtained through the upper 
bound on the temporal change of entropy under the Dirichlet boundary conditions. 
More specifically, the out-flux vector $\vec{j}(\vec{x},t)$ and the quantity $\vec{j}(\vec{x},t)\ln f$ 
are required to vanish at the boundary in combination with the continuity equation 
for probability density
\begin{eqnarray}
\frac{\partial f(\vec{x},t)}{\partial t}+\nabla \cdot\vec{j}(\vec{x},t)=0. 
\end{eqnarray}
One of these expressions was given by Nikolov and Frieden \cite{Niko} as 
\begin{eqnarray}
\frac{dS}{dt}\leqslant \frac{1}{2d}I(t) \frac{d\langle r^2\rangle}{dt},\label{eqn:Niko}
\end{eqnarray}
where $d$ is the spatial dimension and $\langle r^2 \rangle$ the mean square displacement 
of the particle investigated. The fact that the entropy increase rate never reaches infinity 
but is appropriately suppressed by two characteristic quantities offers deep insight. 
That is, it bridges between the information theoretical aspect through the shape of probability 
distribution and the kinematic one through the speed of particle. 
The other one was expressed by Brody and Meister \cite{Brody} as 
\begin{eqnarray}
\frac{dS}{dt}\leqslant \gamma \sqrt{I},\label{eqn:Brody}
\end{eqnarray} 
where coefficient $\gamma$ is given as $\langle (\vec{j}/f)^2\rangle_f$. Moreover, this bound 
represents the collaborative effect of $I$ and the flow of transmitted matter.
We can consider this formula as more versatile than Eq. (\ref{eqn:Niko}), since it gives the bound 
if we specify a physical model (i.e., the form of the flux). On the other hand, a lower bound on $I$ 
itself was alternatively obtained as $\langle \nabla\cdot(\vec{j}/f)\rangle_f^2/\gamma \leqslant I$ 
\cite{Plas2}. Furthermore, according to the Fickian diffusion law, $\vec{j}=-D\nabla f$ with the 
diffusion constant $D$, we can easily obtain $dS/dt=D\cdot I \geqslant 0$. In either case, the quantity 
$I$ is found to play a vital role in limiting the entropy production.\\
 
The present consideration is deeply rooted in the unidirectional characteristics of time (the arrow 
of time). Arrow of time is commonly understood as the consequence of the transformation
of local information into nonlocal correlations, i.e., the production of information measure. 
In this sense, we focus on the rate of change of the Fisher information with time, which is termed as 
''Fisher information production'' $dI/dt$ in analogy to the entropy production $dS/dt$, and we will 
present the upper bound for it in a general setting. The present consideration is of 
high importance not only in terms of the notion of the arrow of time, based on $I$,  
but also from the viewpoint of the second law of thermodynamics. In Sec. II, we derive the lower 
bounds for $I$ with and without the flow and determine the form of the upper bound for $dI/dt$. 
We present examples for these bounds in Sec. III. The lower bounds on the Fisher information production 
are derived in terms of Shannon entropy production via the logarithmic Sobolev inequality in Sec. IV.
Finaly, we discuss the conclusions. 
\section{Novel lower bounds for $I$ and an upper bound for its production}
\subsection{A lower bound expression in presence of flux}
Since the continuity of $\nabla f$ is not obvious from the onset for a given $f$ in general, we 
first need to remark that we assume Leibniz's rule for differentiation under the integral sign; i.e., 
$dI/dt=\int_{\mathbb{R}^1} \frac{\partial}{\partial t}((\nabla f)^2/f) dx$ is assumed to be guaranteed. 
This implies that, in case $x$ and $t$ are in closed intervals, $(\nabla f)^2/f$ is continuous 
for $x$ and $t$. Then, $\frac{\partial}{\partial t}((\nabla f)^2/f)$ is also continuous. 
In case the integral involves an infinite domain, we require 
$\int_{w}^{\infty} \frac{\partial}{\partial t}((\nabla f)^2/f) dx$
to uniformly converge with respect to $t$, where $w$ irrelevant to $t$, such that 
$|\int_{w}^{\infty} \frac{\partial}{\partial t}((\nabla f)^2/f) dx| <\epsilon$, $\forall \epsilon$.\\
In this context, we obtain the time derivative of the one-dimensional Fisher information, which yields
\begin{eqnarray}
\frac{dI}{dt}=\int_{\mathbb{R}^1} \frac{\partial}{\partial t}\left( \frac{(\nabla f)^2}{f}\right)dx=
\int_{\mathbb{R}^1} \left[ \left(\frac{\nabla f}{f}\right)^2\nabla \cdot \vec{j}
-2\frac{\nabla f}{f}\cdot\nabla(\nabla\cdot\vec{j})\right]dx, \label{eqn:dI}
\end{eqnarray}
in which we have used the interchangeability $f_{xt}=f_{tx}$, since we are assuming that both $f_{xt}$ 
and $f_{tx}$ are continuous. In addition, we have substituted the continuity equation 
$\partial_t f=-\nabla \cdot\vec{j}$. Here we continue using the symbol $\nabla$ instead of 
$\partial/\partial x$ because we want to include the high-dimensional case for consideration. 
The fact that $I$ decreases with time and attains its minimum has been shown in several contexts. 
Indeed, the never-increasing property $dI/dt\leqslant 0$ holds true for solutions of the Fokker-Planck 
equation \cite{Plas1}, indicating that the minimum Fisher information (MFI) principle \cite{Fri} 
induces the existence of time asymmetry in terms of $I$ \cite{FS}.  
Accordingly, it is apparently legitimate to expect that in a wider class of time evolution, 
possessing the continuity equation, the Fisher information production is negative. 
In Sec.IV, however, we note a situation, in which this qualification is not satisfied. 
Now, imposing the decreasing $I$, we obtain the following inequality for the 
terms specified in Eq. (\ref{eqn:dI}) 
\begin{eqnarray}
\int_{\mathbb{R}^1} \left[(\nabla \ln f)^2\nabla \cdot \vec{j}\right]dx \leqslant 
2\int_{\mathbb{R}^1} \frac{\nabla f}{\sqrt{f}}\cdot \frac{\nabla(\nabla\cdot \vec{j})}{\sqrt{f}} dx 
\leqslant 2 \sqrt{\int_{\mathbb{R}^1} \frac{(\nabla f)^2}{f}dx} 
\sqrt{\int_{\mathbb{R}^1} \frac{\left[\nabla(\nabla \cdot\vec{j})\right]^2}{f}dx}.
\end{eqnarray}
The last inequality follows from the Schwarz inequality. Therefore, we obtain a lower bound when 
the leftmost term of the above inequality is a positive quantity
\begin{eqnarray}
I \geqslant \frac{1}{4}\frac{\left( \int_{\mathbb{R}^1} dx(\nabla \ln f)^2\nabla \cdot \vec{j}\right)^2}
{\left\langle \left[\frac{\nabla(\nabla \cdot \vec{j})}{f}\right]^2\right\rangle}.\label{eqn:lb}
\end{eqnarray}
This is equivalent to considering only systems with positive divergence flows. We note also that 
the inequality becomes valid when the absolute value of the leftmost part is less than or equal 
to the right hand side, i.e., $|\int_{\mathbb{R}^1} dx(\nabla \ln f)^2\nabla \cdot \vec{j}|\leqslant 
2I\sqrt{\int_{\mathbb{R}^1} dx\left[\nabla(\nabla \cdot\vec{j})\right]^2/f}$.
This condition limits the scope of applicability of the analysed distribution functions, and 
as we shall mention later in Sec. III, the Gaussian distribution lies within this 
applicable class. Consequently, the derived inequality Eq. (\ref{eqn:lb}) is yet another expression 
of a lower bound for $I$, which is different from the previously reported one in \cite{Plas2}. 
In the $d$-dimensional case, the Schwarz inequality gives
\begin{eqnarray}
\int_{\mathbb{R}^d}\left[\frac{\nabla f}{f}\cdot \nabla(\nabla \cdot \vec{j})\right]d\tau \leqslant 
\sum^{d}_{k=1}\left\| \frac{\nabla_k f}{\sqrt{f}}\right\|_2 \left\| \frac{\nabla_k(\nabla \cdot \vec{j})}
{\sqrt{f}}\right\|_2,
\end{eqnarray}
where $d\tau=dx_1\cdots dx_d$. Then, we have the following inequality: 
\begin{eqnarray}
d\mathbf{I}\cdot d\mathbf{J}\geqslant \frac{1}{2}\int_{\mathbb{R}^d}
\left[\left(\frac{\nabla f}{f}\right)^2\nabla \cdot\vec{j}\right]d\tau,
\end{eqnarray}
in which we denote the vector $d\mathbf{I}=((d\mathbf{I})_1,\ldots,(d\mathbf{I})_d)$,  
whose components $\| \nabla_k f/\sqrt{f}\|_2$ ($k=1,\ldots,d$) constitute the Fisher information, such  
that $|d\mathbf{I}|^2=I$. It reflects the shape (geometry) property of the distribution function. 
On the other hand, the vector $d\mathbf{J}$ associated with the 
$L^2$-norm of the quantity $\nabla_k(\nabla \cdot \vec{j})/\sqrt{f}$ provides information on the (phase) 
spatial change of the flow. It is interesting to note that the bound for $dS/dt$ by Brody and Meister 
\cite{Brody} essentially stems from the application of the Schwarz inequality. \\

Now, let us consider the one-dimensional case and evaluate the magnitude $|dI/dt|$, which is 
our second objective. Let us put the absolute values of the rightmost 
integrations in Eq. (\ref{eqn:dI}) respectively as $C_1$ and $C_2$,
\begin{eqnarray}
C_1:=\Big| \int_{\mathbb{R}^1} \frac{\nabla f}{\sqrt{f}}\frac{\nabla f}{\sqrt{f}}
\frac{\nabla \cdot \vec{j}}{f}dx\Big|, \quad 
C_2:=\Big| 2\int_{\mathbb{R}^1}\frac{\nabla f}{f}\cdot\nabla(\nabla \cdot \vec{j})dx\Big|.
\end{eqnarray}
We apply the H{\"o}lder inequality for the three functions $f_1$, $f_2$, and $f_3$
\begin{eqnarray}
\Big|\int_{\mathbb{R}^1} dx f_1f_2f_3\Big| \leqslant \|f_1\|_{p_1}\|f_2\|_{p_2}\|f_3|_{p_3}\quad  
{\rm with} \quad \frac{1}{p_1}+\frac{1}{p_2}+\frac{1}{p_3}=1,
\end{eqnarray}
where $\|f\|_p(t)=(\int_{\mathbb{R}^1} |f(x,t)|^p dx)^{1/p}$ ($1\leqslant p <\infty, t>0$) is the $L^p$ norm. 
Then, by setting $p_1=p_2=2$ and $p_3=\infty$, the limit for $C_1$ can be expressed as follows
\begin{eqnarray}
C_1 \leqslant \left\| \frac{\nabla f}{\sqrt{f}}\right\|_2^2 
\left\| \frac{\nabla \cdot \vec{j}}{f}\right\|_{L^\infty},
\end{eqnarray}
where $\| \cdot\|_{L^\infty}=ess.\sup|\cdot|$. 
In conjunction with the triangle inequality for the last expression of Eq. (\ref{eqn:dI}), 
we obtain the upper bound 
\begin{eqnarray}
\Big|\frac{dI}{dt}\Big| \leqslant C_1+C_2 = \alpha I + \beta\sqrt{I}, \label{eqn:dIdt}
\end{eqnarray}
where we have put the coefficients respectively as
\begin{eqnarray}
\alpha=\mathop{ess.\sup}\limits_{x\in \mathbb{R}^1}
\Big| \frac{\nabla \cdot \vec{j}}{f}\Big| \quad {\rm and} \quad   
\beta= 2\sqrt{\int_{\mathbb{R}^1}\frac{[\nabla(\nabla\cdot\vec{j})]^2}{f}dx}.\label{eqn:ab}
\end{eqnarray}
\subsection{Lower bound expressions in terms of the Sobolev and logarithmic Sobolev inequalities}
In this section, we provide the lower bounds for $I$ depending on dimensions, when we do not consider  
the flux. Since the Fisher information captures the coarse-grained inclination 
of the distribution function, it is useful to find the operative limit of the average gradient 
in terms of the spreading of the function. \\

For a function $g$ that vanishes at infinity with its gradient in $L^2(\mathbb{R}^n)$, i.e.,  
$g\in D^1(\mathbb{R}^n)$, the Sobolev inequality with dimension $n\geqslant 3$ (e.g. \cite{Lieb}) 
reads 
\begin{eqnarray}
\left\| \nabla g\right\|_2^2 \geqslant \frac{n(n-2)}{4}|\mathbb{S}^n|^\frac{2}{n}
\left\| g\right\|_{\frac{2n}{n-2}}^2,
\end{eqnarray}
where $|\mathbb{S}^{n-1}|=2\pi^{\frac{n}{2}}/\Gamma(n/2)$ is a sphere of radius $1$ in $\mathbb{R}^n$. 
By substituting $g=\sqrt{f}$ we obtain 
\begin{eqnarray}
I(f)\geqslant \frac{n(n-2)2^{\frac{2}{n}}\pi^{1+\frac{1}{n}}}{\Gamma\left(\frac{n+1}{2}\right)^\frac{2}{n}}
\left\| f\right\|_{\frac{n}{n-2}}.
\end{eqnarray}
In case of $n=2$, for function $g$ and its gradient in $L^2(\mathbb{R}^2)$, the following inequality holds 
\begin{eqnarray}
\left\| \nabla g\right\|_{H^1(\mathbb{R}^2)}^2 \geqslant C\left\| g\right\|_{L^q(\mathbb{R}^2)}^2, \quad 
(2\leqslant q < \infty)
\end{eqnarray}
where the norm of the Sobolev space $H^1(\mathbb{R}^2)$ is defined as
\begin{eqnarray}
\left\| \nabla g\right\|_{H^1(\mathbb{R}^2)}:=\left( \int_{\mathbb{R}^2}|g(x)|^2 dx + 
\int_{\mathbb{R}^2}|\nabla g(x)|^2 dx \right)^{1/2},
\end{eqnarray}
and constant $C$ depends only on $q$. Then, noting that $\left\| f\right\|_2^2=1$ by the normalization 
of the probability function, we obtain the lower bound 
\begin{eqnarray}
I(f)\geqslant 4(C\left\| f\right\|_{\frac{q}{2}}-1).
\end{eqnarray}
In case of $n=1$, for any $g\in H^1(\mathbb{R}^1)$,
\begin{eqnarray}
\left\| \frac{dg}{dx}\right\|_2^2+\left\| g \right\|_2^2 \geqslant 2\left\| g \right\|_\infty^2
\end{eqnarray}
holds. By setting $g=\sqrt{f}$, we have the inequality 
\begin{eqnarray}
I(f)\geqslant 4(2\left\| \sqrt{f} \right\|_\infty^2-1).\label{eqn:sb1}
\end{eqnarray}
We use this inequality in Sec. III.
\section{Examples}
\subsection{Normal diffusion}
As the benchmark evaluation of the bounds derived here, we examine the one-dimensional 
Wiener process, where the Gaussian probability distribution function with the time-dependent 
dispersion $\sigma(t)$ and with the associated flow satisfying the continuity equation 
govern the system. This example has also been previously considered in the literature \cite{Brody,Plas2}. 
The flow and the distribution function are related as $\vec{j}(x,t)=xf\dot{\sigma}(t)/\sigma(t)$ 
according to the continuity equation and the relation $j=-\partial f/\partial x$.
By straightforward calculations, we obtain
\begin{eqnarray}
\left\langle \left(\frac{j_{xx}}{f}\right)^2\right\rangle=\frac{6\dot{\sigma}^2(t)}{\sigma^4(t)}, 
\quad \int^{\infty}_{-\infty}(\partial_x\ln f)^2 j_xdx=-\frac{2\dot{\sigma}(t)}{\sigma^3(t)}.
\end{eqnarray}
Since the Fisher information is calculated to be $I=1/\sigma^2(t)$, we can corroborate that the 
applicability condition of Eq. (\ref{eqn:lb}) is indeed satisfied, 
$|-2\dot{\sigma}(t)/\sigma^3(t)|<2\sqrt{I}\sqrt{\langle \left(j_{xx}/f\right)^2\rangle}$. 
Therefore, it is well-justified to insert these into Eq. (\ref{eqn:lb}), and thus, we obtain the 
lower bound 
\begin{eqnarray}
I\geqslant \frac{1}{6\sigma^2(t)}.
\end{eqnarray}
This provides a tighter bound compared to $\sigma^{-2}(t)$ as the lower bound obtained in \cite{Plas2}  
from the derived inequality (i.e., $I \geqslant \langle \nabla\cdot(\vec{j}/f)\rangle_f^2/\gamma$) therein. 
A possible interpretation of the origin of tightness is that the occurrence of flow reduces 
the information in a system compared to the case without it. In this sense, the fact that the value 
of the lower bound $\sigma^{-2}(t)$ \cite{Plas2} coincides with that of the Fisher information $I$,  
determined only from the form of the distribution, does not convey limitation, at least for 
one-dimensional normal diffusion, and the present lower bound may be a preferable alternative. 
The coefficient $1/6$ is for the one-dimensional case and it differs for the 
other dimensions. Moreover, we note that the lower bound Eq. (\ref{eqn:sb1}), derived from the 
Sobolev inequality, for one dimension is indeed found to be satisfied, by considering  
\begin{eqnarray}
\left\| \sqrt{f} \right\|_\infty^2=(\mathop{ess.\sup}\limits_{x\in \mathbb{R}^1}|\sqrt{f}|)^2
=\frac{1}{\sqrt{2\pi}\sigma}
\end{eqnarray}
and by the fact $(\sigma-1/\sqrt{2\pi})^2+1/4-1/2\pi\geqslant 0$.
\subsection{Truncated Gaussian}
The use of truncated Gaussian distributions is common in statistics, econometrics, and 
in many other areas of science, where the probability density function has a cutoff from below or 
above (or both) while keeping the Gaussian form (e.g., \cite{Greene}). The density function of the 
truncated normal distribution defined in $x\in [a,b]$ is 
\begin{eqnarray}
\frac{\frac{1}{\sigma}\phi(\frac{x-\mu}{\sigma})}{\Phi(\frac{b-\mu}{\sigma})-\Phi(\frac{a-\mu}{\sigma})}, 
\end{eqnarray}
where $\phi(x)$ is the standard normal distribution with mean $\mu$ and variance $\sigma^2$.  
$\Phi(x)$ denotes its cumulative distribution. Below, we use a time-dependent finite interval 
$x\in [-\sigma(t), \sigma(t)]$ by setting $\mu=0$ and proceed without the scaling 
factor $\sigma(t)(\Phi(1)-\Phi(-1))$, derived from the normalization. Then, the Fisher information 
is calculated to be 
\begin{eqnarray}
I=\int^{\sigma(t)}_{-\sigma(t)}\frac{(f_x)^2}{f}dx =\frac{1}{\sigma^2(t)}
\left\{ \rm{Erf}\left( \frac{1}{\sqrt{2}}\right)-\sqrt{\frac{2}{\pi e}}\right\},
\end{eqnarray}
where the error function $\rm{Erf}(1/\sqrt{2})=0.6826...$ and the positivity $I>0$ is accordingly 
maintained. The use of the truncated Gaussian in our consideration is equivalent to incorporation  
of the flow form, same as that given in the previous example, because during the process, 
the distribution keeps the Gaussian within finite support. 
For the coefficients $\alpha$ and $\beta$ defined in Eq. (\ref{eqn:ab}), we calculate respectively as  
\begin{eqnarray}
\alpha=\mathop{ess.\sup}\limits_{x\in [-\sigma(t),\sigma(t)]}
\Big| \frac{j_x}{f}\Big|=\frac{\dot{\sigma}(t)}{\sigma(t)},\quad
\beta = 2\sqrt{\int^{\sigma(t)}_{-\sigma(t)}\frac{(j_{xx})^2}{f}dx}
=\frac{12\dot{\sigma}^2(t)}{\sigma^4(t)}\left\{ \rm{Erf}\left( \frac{1}{\sqrt{2}}\right)
-\sqrt{\frac{2}{\pi e}}\right\}.
\end{eqnarray}
Therefore, we find that the absolute value of the time derivative is bounded from above as 
\begin{eqnarray}
\Big|\frac{dI}{dt}\Big| \leqslant \frac{\dot{\sigma}(t)}{\sigma(t)} I
+12\frac{\dot{\sigma}^2(t)}{\sigma^4(t)} I^{3/2}.\label{eqn:trG}
\end{eqnarray}
Since we are dealing with moving boundaries (truncation positions) here, the time derivative of the 
integral of a bivariable function, whose limits depend on time has additional terms (e.g., \cite{Takagi}) as 
\begin{eqnarray}
\frac{d}{dt}\int^{u(t)}_{v(t)}\rho(x,t)dx=\dot{u}(t)\rho(u(t),t)-\dot{v}(t)\rho(v(t),t)+
\int^{u(t)}_{v(t)}\frac{\partial}{\partial t}\rho(x,t)dx,
\end{eqnarray}
where $\rho(x,t)=-f(x,t)\log f(x,t)$ in our case and $\int^{u(t)}_{v(t)}\rho(x,t)dx$ is assumed to be 
continuous within the interval of interest in both $x$ and $t$. We note that the first two extra terms 
calculated as $\dot{\sigma}(t)\rho(\sigma(t),t)-(-\dot{\sigma}(t))\rho(-\sigma(t),t)$ contribute 
only in shifting the upper bound, obtained in Eq. (\ref{eqn:trG}). Substituting the Gaussian form of 
$f$ into $\rho$, we have $\rho(\sigma(t),t)=\sigma^{-1}(t)(1/2+\log\sqrt{2\pi}+\log\sigma(t))/\sqrt{2\pi e}$.
Then, the final upper bound becomes
\begin{eqnarray}
\Big|\frac{dI}{dt}\Big|\leqslant \frac{\dot{\sigma}(t)}{\sigma(t)} I+12\frac{\dot{\sigma}^2(t)}{\sigma^4(t)} I^{3/2}
+ \frac{\dot{\sigma}(t)}{\sigma(t)\sqrt{2\pi e}}\left( 1+\log(2\pi)+2\log\sigma(t)\right).\label{eqn:mvb}
\end{eqnarray}
Note that by comparing Eq. (\ref{eqn:mvb}) with Eq. (\ref{eqn:trG}), the effect of moving boundary 
appears as an increment in the bound.
\section{A bound for $dI/dt$ in terms of entropy production}
In Eqs. (\ref{eqn:Niko}) and (\ref{eqn:Brody}), viewed conversely, $I$ and its square root 
are respectively bounded from below by entropy production $dS/dt$. Therefore, we can conceive  
the available bound in any form to generate change in Fisher information in terms of flow.
Specifically, we develop an interest in obtaining a general expression of the lower bound for the 
Fisher information production $dI/dt$ in terms of the entropy production when the systems follow 
the continuity equation $\partial_t f=-\nabla\cdot \vec{j}$ with the aid of the interdependence between  
$S$ and $I$. One may think that finding the lower bound on $dI/dt$ contradicts the arrow of time  
employed in Sec. II. However, this characterization can only be true under 
well-organized circumstances such as systems in which the model follows the heat equation, 
although it is widely expected to hold as mentioned in Sec. II A. In fact, Fisher 
information does not necessarily decrease in time in certain cases. Indeed, within properly 
developing living cells, $I$ is locally maximized, i.e., within the cell delimited by the 
membrane \cite{bio}. This can occur at the expense of transferring increased waste and disorder 
outside the cell. The latter ensures that the second law of thermodynamics is overall obeyed. 
Therefore, we proceed as follows.\\ 
Recalling that the logarithmic Sobolev inequality for the function $g$ (e.g., \cite{Lieb})
\begin{eqnarray}
\frac{1}{\pi}\int_{\mathbb{R}^n} |\nabla g(x)|^2 dx\geqslant \int_{\mathbb{R}^n} |g(x)|^2 
\ln\left( \frac{|g(x)|^2}{\left\| g \right\|_2^2}\right)dx + n\left\| g \right\|_2^2.
\end{eqnarray}
Setting $g=\sqrt{f}$ and by normalization of the probability density, we find 
\begin{eqnarray}
I(f)\geqslant 4\pi(n-S(f)),\label{eqn:lsi}
\end{eqnarray}
where the Shannon entropy is $S(f)=-\int_{\mathbb{R}^n}f(x)\ln f(x)dx$.
Unless the function $I(f)+4\pi S(f)$ is a monotonically decreasing function with respect to time, 
there should exist at least a time region where its derivative becomes positive. 
Recalling that the identity $dS/dt=\langle {\rm div}(\vec{j}/f)\rangle_f$ holds true \cite{Plas2} 
under the setting same as ours, we can rewrite it further as 
\begin{eqnarray}
\frac{dS}{dt}= \left\langle \frac{{\rm div}\vec{j}}{f}\right\rangle_f 
-\left\langle \vec{j}\cdot \frac{\nabla f}{f^2}\right\rangle_f,
\end{eqnarray}
where we have used the formula ${\rm div}(\phi \vec{A})=\vec{A}\cdot \nabla \phi+\phi\nabla\cdot \vec{A}$ 
for a scalar function $\phi$ and a vector $\vec{A}$. 
Therefore, by differentiating Eq. (\ref{eqn:lsi}) we obtain the following lower bound 
\begin{eqnarray}
\frac{dI}{dt} \geqslant 4\pi\left( \left\langle \vec{j}\cdot \frac{\nabla f}{f^2}\right\rangle_f 
- \left\langle \frac{{\rm div}\vec{j}}{f}\right\rangle_f \right).
\end{eqnarray}
A direct evidence for the contradiction between the property $dI/dt\leqslant 0$ and the above-derived 
inequality can be checked in case of one-dimensional normal diffusion. That is, we obtain 
$dI/dt\geqslant 4\pi\dot{\sigma}/\sigma$, which is finite as long as particles diffuse 
(i.e., $\dot{\sigma}>0$). This fact shows the scope of the applicability.
\section{Discussion}
If we specify the relation between flux vector $\vec{j}$ and distribution $f$, we obtain a physical model.
In general, we can write the process as $P(L)f(x,t)=j(x,t)$ in the one-dimensional case, where $P(L)$ 
is a linear operator, represented by a polynomial of the differential operator $L=\partial/\partial x$ 
with constant coefficients. Fick's law is a special case, which follows from the choice $P(L)=-DL$. 
Since the Fourier transform of the both sides becomes $P(\xi)\hat{f}(\xi)=\hat{j}(\xi)$ when 
$P(\xi)\neq 0, (\xi \in \mathbb{R}^1)$, we can have the expression of the original distribution 
$f(x,t)=(\sqrt{2\pi})^{-1}\int_{\mathbb{R}^1}d\xi e^{ix\xi}\hat{j}(\xi)/P(\xi)$ by inverse transform. 
Alternatively, in terms of the inverse operator $P^{-1}(L)$, we have the expression 
$f(x,t)=P^{-1}(L)j(x,t)$. Together with the definition of the Fisher information in Eq. (\ref{eqn:Idef}), 
we regard $I$ as the result of the averaging function $\chi(f)$ with $f$ determined, such that
\begin{eqnarray}
I(t)=\langle \chi(f)\rangle_f
\end{eqnarray}
holds. If we specify the functional form of $\chi(f)$ as $\chi(f)=(j/f)^2$, it is equivalent to choose 
the form of flux as Fick's law $j=-\partial f /\partial x$. Other models are realized by fixing 
the form $\chi$ as a function of flux and distribution. Assuming that $\partial \chi/\partial t$ 
is continuous, we have 
\begin{eqnarray}
\frac{dI}{dt}=\int_{\mathbb{R}^1}\left( \frac{\partial f}{\partial t}\chi 
+ f \frac{\partial \chi}{\partial t}\right)dx.
\end{eqnarray}
The upper bound of the above equation relies on the supremum, which is derived from the two competing terms 
$(\partial_t{f})\chi$ and $f\partial_t{\chi}$, but the Schwarz' inequality at least provides the maxima for 
each term. Then, we have 
\begin{eqnarray}
\Big|\frac{dI}{dt}\Big| \leqslant \left\| \frac{\partial f}{\partial t}\right\|_2 \left\| \chi \right\|_2
+\left\| f\right\|_2 \left\| \frac{\partial \chi}{\partial t}\right\|_2, 
\end{eqnarray}
which is a desirable expression for a bound with $L^2$ norms because we normally require square 
integrability of distributions in physics.
\section{Conclusion} 
Apart from the well-known Cram{\'e}r-Rao bound \cite{CR}, which asserts that $I$ cannot exceed the 
inverse of the mean square error of a measured quantity, the fact that the bounds for $I$, $|dI/dt|$,  
and $dI/dt$, derived from the distribution functions, obey the laws of physics definitely links physics and 
the information contained in a physical system. The former (Cram{\'e}r-Rao) originates from repeated active 
measurements (i.e., statistics), but in contrast, the latter originates from the flux $\vec{j}$ of a 
physical entity. In this context, we have presented new alternative upper and lower bounds for the time 
derivative of $I$. However, we neither have nor established a relation between the information flux 
and information production for $I$, whereas in nonequilibrium thermodynamics \cite{deG} based on 
Shannon entropy, there is a familiar local formulation, i.e., $dS/dt= -\nabla\cdot \mathbf{J}+\sigma$, 
where $\mathbf{J}$ and $\sigma$ denote the entropy flux and the entropy production, respectively. 
In irreversible processes, $\sigma\geqslant 0$ implies the Boltzmann's H-theorem. 
A search for the counterpart may lead to a deeper understanding of the informational structures, inherent 
in physical systems.
\section{Acknowledgement}
The author wishes to thank the reviewer and the editor for valuable comments and suggestions
to improve the presentation of the paper.
 
\end{document}